\documentclass[12pt]{article}

 \usepackage{hyperref}
\hypersetup {
pdfpagemode = {UseNone},
pdftitle = {Space-Time-Matter},
pdfauthor = {Christian Beck},
pdflang = {en-GB}
}

\makeatother 
\renewcommand{\vector}[1]{\ensuremath{\boldsymbol{#1}}} 

\newcommand{\braket}[2]{\left< #1 \vphantom{#2} \right|
 \left. #2 \vphantom{#1} \right>} 
 \newcommand{\tra}[2]{ \operatorname{Tr}_{\mathcal{H}_{#1}}  \left[ #2  \right]}

   \newcommand{\prob}[1]{\mathbb{P}^{\psi}\left( #1\right)}

    \newcommand{\D}{\mathscr{D}}
      \newcommand{\Dtd}{\mathscr{D}_{(t,\Delta)}}
      \newcommand{\Dtdp}{\mathscr{D}_{(t,\Delta')}}

\let\baraccent=\= 
\renewcommand{\=}[1]{\stackrel{#1}{=}} 

\newcommand{\xcentcolon}{%
  \mathrel{\vbox{\hbox{$:$}\kern.2ex}}%
}

\usepackage{dsfont,a4wide,amsthm,amscd,mathrsfs,amsmath,graphicx,floatflt,slashed,txfonts}

\usepackage[noblocks]{authblk}

    \usepackage{filecontents}

\usepackage{bbm}

\usepackage{tabularx}
\usepackage[]{units}

\usepackage{verbatim}

    {\cleardoublepage\null\vfill\begin{center}%
    \bfseries Acknowledgement\end{center}}%
    {\vfill\null}

\makeatletter
\def\oversortoftilde#1{\mathop{\vbox{\m@th\ialign{##\crcr\noalign{\kern3\p@}%
      \sortoftildefill\crcr\noalign{\kern3\p@\nointerlineskip}%
      $\hfil\displaystyle{#1}\hfil$\crcr}}}\limits}

\def\sortoftildefill{$\m@th \setbox\z@\hbox{$\braceld$}%
  \braceld\leaders\vrule \@height\ht\z@ \@depth\z@\hfill\braceru$}

\makeatother

\usepackage[format=plain,labelsep=newline,font=sl,textfont=small,labelfont=sc,justification=centerlast]{caption}
\usepackage{amsmath}
\usepackage{amsfonts}
\usepackage{amssymb}
\usepackage{stmaryrd}

\usepackage{fancyhdr}
\fancypagestyle{plain}{
}

\usepackage{emptypage}

\usepackage{shadethm,thmbox,color}

\newlength\thmvorhernachher
\makeatletter
\let\thmbox@head@org\thmbox@head
\let\thmbox@tail@org\thmbox@tail
\renewcommand*\thmbox@head{%
\vspace{\thmvorhernachher}
\thmbox@head@org
}
\renewcommand*\thmbox@tail{%
\thmbox@tail@org
\vspace{\thmvorhernachher}
}
\makeatother

\newtheorem{thm}{Theorem}

\fancyhfoffset{0.0pt}

\newcommand{\supp}{\mathop{\mathrm{supp}}}

\DeclareMathOperator{\tr}{Tr}

\newcommand{\N}{\mathbb{N}}

\newcommand{\C}{\mathbb{C}}
\newcommand{\R}{\mathds{R}}

\def\ben{\begin{equation*}}
\def\een{\end{equation*}}  

\def\be{\begin{equation}}
\def\ee{\end{equation}}  

\def\bg{\begin{equation} \begin{gathered}}
\def\eg{\end{gathered} \end{equation}}  
 
 \def\bgn{\begin{equation*} \begin{gathered}}
\def\egn{\end{gathered} \end{equation*}}  

\def\bs{\begin{equation} \begin{split}}
\def\es{\end{split} \end{equation}}

\def\bsn{\begin{equation*} \begin{split}}
\def\esn{\end{split*} \end{equation}}

\title{ \vspace*{-1cm} \Huge{{\scshape {Space--Time--Matter} \\ \vspace{.2cm} \Large {Some Notes on the Localization Problem \vspace{-.4cm} \\  in Relativistic Quantum Theory}}}} 
\author[*]{Christian Beck\thanks{christian\_beck@posteo.de}}
\date{\vspace*{-.5cm}May 2023}

\begin{document}

 \maketitle
\thispagestyle{empty} 

 \vspace*{-1cm}\begin{abstract}

 This work aims to shed some light on the meaning of the positive energy assumption in relativistic quantum theory and its relation to questions of localization of quantum systems. It is shown that the positive energy property of solutions of relativistic wave equations (such as the Dirac equation) is very fragile with respect to state transformations beyond free time evolution. Paying attention to the connection between negative energy Dirac wave functions and pair creation processes in second quantization, this analysis leads to a better understanding of a class of problems known as the localization problem of relativistic quantum theory (associated for instance with famous results of Newton \& Wigner, Reeh \& Schlieder, Hegerfeldt or Malament). Finally, this analysis is reflected from the perspective of a Bohmian quantum field theory.

 \end{abstract}

  \tableofcontents

\newpage

\setcounter{page}{1}

\section{A Basic Theorem} \label{basthmsect}

We start with a result that follows from complex analysis of several complex variables:\vspace{-.5cm}

\begin{thm} \label{basres}  
Let $\lambda$ be a complex measure\footnotemark on $\R^4$ with support in the closure of the forward light cone $\overline{V}_+=\left\{p\in\R^4\mid p_{\mu}p^{\mu}=p_0^2-\vector{p}^2\geq 0,\;p_0\geq 0\right\}$ of the origin. Consider the function $f:\R^4\to\C$ given by
\be \label{fanacont}
f(x)= \int  \; e^{i  p x} \; d^4\lambda(p) 
\ee
where $px=p_{\mu} x^{\mu}$ is the Minkowski scalar product. If $f$ vanishes on an open connected subset $\mathcal{O}\subset\R^4$ it follows that $f\equiv 0$ on all of $\R^4$. 
\end{thm} \vspace{-.5cm}
\footnotetext{A complex measure can be always understood as a collection of four ordinary measures, it has a real and an imaginary part which are signed measures. These in turn can each be decomposed into two normal finite measures using a Hahn-Jordan decomposition. The important thing here about a complex measure is that it is always finite (e.g. a finite ordinary measure is also a complex measure). That $\lambda$ has support in $\overline{V}_+$ means that all integrals with respect to $\lambda$ over subsets of $\R^4$ disjoint from $\overline{V}_+$ vanish, in particular $\int_{\R^4}d^4\lambda(p)=\int_{\overline{V}_+}d^4\lambda(p)\in \C$.}

The proof can be found in \cite{MyBook} (see corollary 4.6). It is based on the fact that $f$ can be continued analytically to a region of $\C^4$ which has $\R^4$ as a part of its boundary. Thus, $f$ in \eqref{fanacont} can be regarded as the boundary value of an analytic function and the conclusion of theorem \ref{basres} then follows with the help of generalizations of the Schwartz reflection principle and the identity theorem to functions of several complex variables.

Theorem \ref{basres} has a number of strong physical consequences for (relativistic) quantum mechanics, all of which are related in some sense and some of which will be discussed in this work. Physically $x$ corresponds to a spacetime vector and $p$ to the energy momentum four-vector. The condition $p\in\overline{V}_+$ is the so-called \textit{spectrum condition}. It is adapted to relativistic considerations and says that the relativistic energy $p_0$ is positive in every Lorentz frame. 

\section{Implications for Wave Functions}

In this section we think of $f$ as (a component of) a relativistic wave function of positive energy, e.g., a positive energy solution of the free Klein-Gordon equation or a spinor component of a positive energy solution of the free Dirac equation. Such functions can be written in the form (see, e.g., \cite{Schweber, schwablAQM})
\be \label{wfform}
\psi(\vector{x},t)=\int \; e^{i (p_0 t-\vector{p}\cdot\vector{x})}\, \delta\left(p_0^2-(\vector{p}^2+m^2)\right)\, \theta(p_0)\, \hat{\psi}({p})\, d^4p
\ee
 which is of the form \eqref{fanacont} with the complex measure\footnote{To be precise, $\int \delta(p^2-m^2)\, \theta(p_0)\, \hat{\psi}({p}) \, dp_0$ must be in $L^1(\R^3,d^3p)$ to define a complex measure.} $d^4\lambda(p) \,=\, \delta(p^2-m^2)\, \theta(p_0)\, \hat{\psi}({p})\,  d^4p$ ($\theta$ denotes the Heaviside step function).  We shall switch in the following between the notations $\psi(\vector{x},t)=\psi_t(\vector{x})=\psi(x)$ (with $x\in\R^4)$, depending on which is most appropriate for the current purpose. 

\subsection{Causally Propagating Positive Energy Wave Functions cannot vanish in a Region}\label{notvanish}

\begin{figure}
    \centering
    {\includegraphics[scale=1.1]{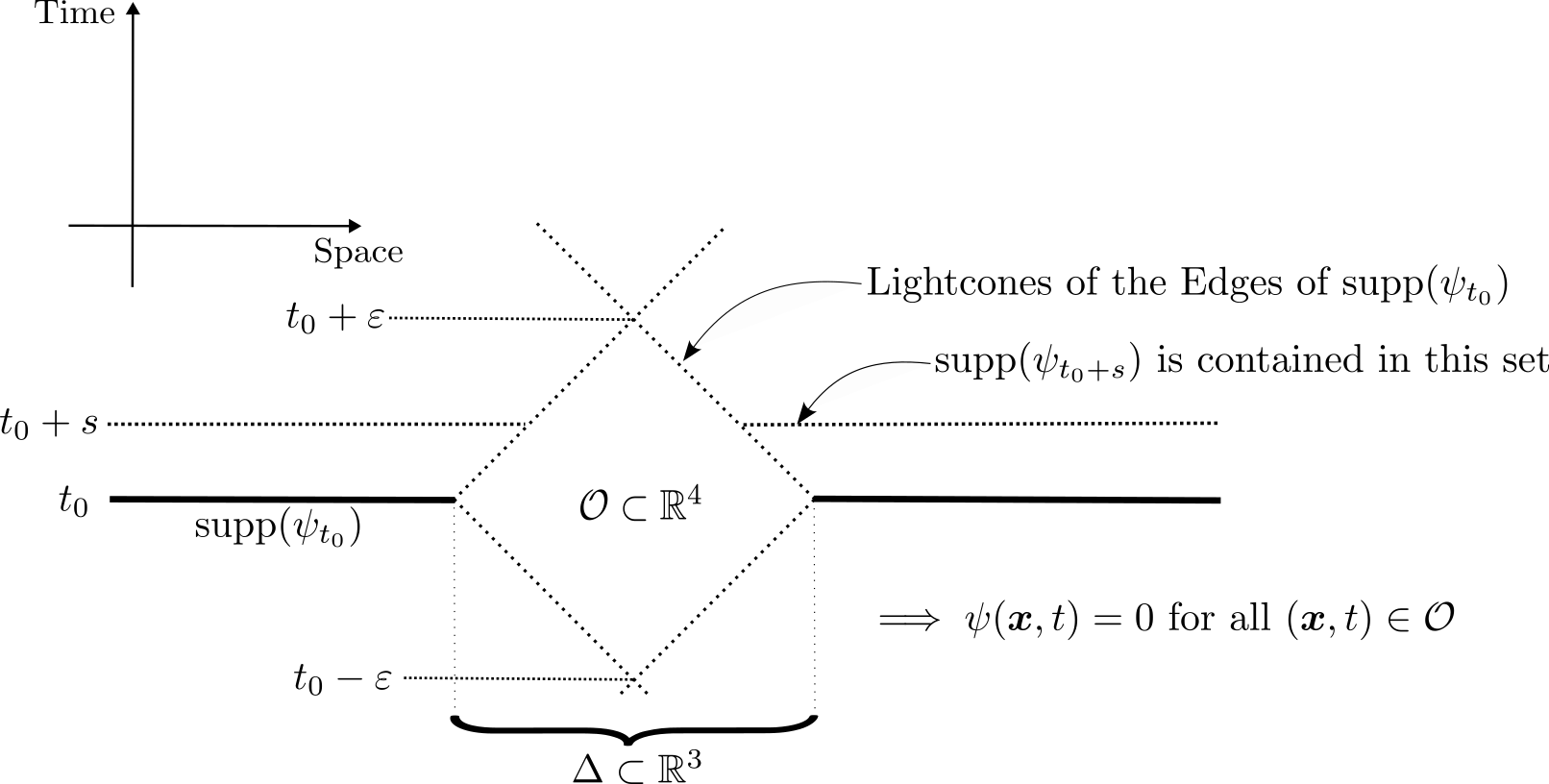}}
    \caption{Causal propagation of the support of a relativistic wave function: if the support of $\psi_t$ can propagate at most at the speed of light and $\psi_{t_0}(\vector{x})=0$ for all $\vector{x}$ in a connected open spatial set $\Delta\subset\R^3$, then $\psi_{t}(\vector{x})$ also vanishes for $(t,\vector{x})$ in a connected open space-time set $\mathcal{O}\subset\R^4$ (the interior of the diamond in the middle). The diagonal dotted lines depict (the essential parts of) the forward and backward light cones of the edges of the support of $\psi_{t_0}$.}
    \label{suppPsi}
\end{figure}

Now suppose $\psi_t(\vector{x})$ vanishes at some time $t=t_0$ on an open, connected spatial subset (region) $\Delta\subset\R^3$, i.e., $\psi_{t_0}(\vector{x})=0$ for all $x\in\Delta$. If $\psi$ propagates causally (which is the case for solutions of relativistic wave equations because of their hyperbolic form \cite{hyperbolPDE, thallerbook}), for later (and earlier) $t$ the support of $\psi_t$ can spread at most with the speed of light as $t$ evolves. Therefore, there must be an $\varepsilon>0$ such that for each $s\in(-\varepsilon, \varepsilon)$, $\psi_{t_0+s}(\vector{x})=0$ on an open spatial set $\Delta_s\subset\R^3$. This way $\psi(x)=0$ for all $x=(t,\vector{x})$ in an open subset $\mathcal{O}\subset\R^4$ (see Fig. \ref{suppPsi}, where the sets $\Delta_s$ are not depicted, but the dashed line at $t_0+s$ indicates the complement of $\Delta_s$). Theorem \ref{basres} thus entails that $\psi_t(\vector{x})=0$ for all $\vector{x}$ and $t$ which contradicts the assumption that $\psi$ is a wave function.

The conclusion is that a causally propagating wave function of the form \eqref{wfform} has at each time the property 
\be 
\supp(\psi)=\R^3
\ee 
This implies in particular the often quoted statement that \textit{relativistic wave functions of positive energy cannot have compact support} but have always \textit{infinite tails}.

It is interesting to note that an analogous statement can also be made for non-relativistic Schr\"odinger wave functions. Theorem \ref{basres} has been formulated in a way that is well suited for relativistic analysis. However, a result analogous to theorem \ref{basres} can be proved \cite{borcherslemma}, which instead of the spectrum condition (that the four momentum $p$ vanishes outside of its forward light cone) only needs the condition that the Hamiltonian (the generator of time translations), whose eigenvalues correspond to the allowed values of $p_0$ in \eqref{wfform}, is bounded from below. This is true in particular for the Schr\"odinger Hamiltonian of non-relativistic quantum mechanics. Since Schr\"odinger wave functions can be zero on open connected sets (as Dirac wave functions can if contributions from negative energy eigenstates are allowed), this shows that Schr\"odinger wave functions spread instantaneously (with infinite propagation velocity) under the free time evolution. 

In a sense, these interrelations can be seen as the core of \textit{Hegerfeldt's theorem}\footnote{Hegerfeldt's theorem proves, roughly said, instantaneous spreading of any `localization probabilities' in quantum theory (with Hilbert space $\mathcal{H}$) with positive energy, if there is a bounded spatial region $\Delta\subset\R^3$ and $\psi\in\mathcal{H}$, such that $\prob{\Delta}=1$ (perfect localization). The probabilities are assumed to be given by the quantum formalism, i.e., for any spatial region $\Delta\subset\R^3$ there is a positive bounded operator $D_{\Delta}$, such that $\prob{\Delta}=\braket{\psi}{D_{\Delta}\,\psi}$. The connection to our discussion above becomes apparent when we choose the PVM of the standard position operator (indicator functions in position representation) $D_{\Delta}=\chi_{\Delta}$ and observe that $1=\|\psi\|^2=\int_{\R^3} |\psi(\vector{x})|^2 d^3x=\int_{\Delta} |\psi(\vector{x})|^2 d^3x=\braket{\psi}{\chi_{\Delta}\,\psi}$ implies that $\psi(\vector{x})=0$ almost everywhere in the complement of $\Delta$. For this choice, Hegerfeldt's theorem thus states that a compactly supported positive energy wave function cannot propagate causally. Hegerfeldt's theorem can be proven by application of theorem \ref{basres} with the choice $f(x)=\braket{\psi}{U(x)\,\psi}$, where $U(x)$ is a unitary representation of space-time translations (see \cite{MyBook}).} \cite{hegerfeldt1985, hegerfeldt1998a, hegerfeldt1980}.

 \vspace{.5cm}

\subsection{A Causally Propagating Positive Energy Wave Function is Completely Determined by its Values in any Region} 

Consider two wave functions $\psi$ and $\psi'$ of the form \eqref{wfform} and suppose that at a certain time $t_0$ there exists an (arbitrarily small) open connected spatial set $\Delta\subset\R^3$ on which the wave functions coincide:
\be 
\psi(\vector{x},t_0)=\psi'(\vector{x},t_0) \quad \text{ for all }  x\in\Delta 
\ee 
Together with $\psi$ and $\psi'$, the wave function 
\be 
\Phi(\vector{x},t):=\psi(\vector{x},t_0)-\psi'(\vector{x},t_0)
\ee 
is also of the form \eqref{wfform}. However, at time $t_0$, $\Phi$ obviously vanishes on $\Delta$ so that theorem \ref{basres} together with our discussion in section \ref{notvanish} proves that $\Phi$ (and thereby either $\psi$ or $\psi'$ or both) cannot propagate causally. The other way around, this entails that two positive energy solutions of relativistic wave equations--which always propagate causally--cannot coincide on any open connected spatial set. 

\vspace{.5cm}

\subsection{Transformations of Causally Propagating Positive Energy Wave Functions are very Special (e.g. necessarily Nonlocal)}

Consider a solution $\psi$ of a relativistic wave equation which is exposed to a local potential $\phi$ for some time, resulting in a transformed state $U_{\phi}\,\psi$, where $U_{\phi}$ is the unitary time evolution with potential $\phi$. Let $U$ be the free time evolution without potential corresponding to the same period of time as $U_{\phi}$. Since the local potential can only locally perturb the wave function and since solutions of relativistic wave equations propagate causally, $U\,\psi$ and $U_{\phi}\,\psi$ can also differ only locally, i.e. $\Phi=U\,\psi-U_{\phi}\,\psi$ has compact support and thus cannot be a positive energy solution. Consequently, if $U\,\psi$ is a positive energy solution (which is the case if $\psi$ has positive energy since the free time evolution leaves the positive energy property invariant), $U_{\phi}\,\psi$ must have contributions from the negative energy spectrum. 

We can also formally set $U=1$ to see that any local transformation of a relativistic positive energy state destroys its positive energy property. In other words, if we wiggle such a wave function just a little bit in the neighborhood of some point, immediately the whole function must change in a non-trivial way, if the resulting wave function shall continue to have positive energy. So a relativistic time evolution cannot be of this kind, it must either act on the wave function on the whole space (including the tails) in a very special way (as free time evolution does) or violate the positive energy property. Note the emphasis on `very special': since the whole function is completely determined by its values in an arbitrarily small neighborhood, its global transformation must be perfectly concerted across all regions if it shall preserve the positive energy property!

\subsection{Discussion} \label{dissect1}

\noindent \textbf{Tails:} The infinite tails of positive energy wave functions do not contradict the fact that positive energy wave functions usually are, for all practical purposes, perfectly localized in bounded spatial regions. Various localization schemes for positive energy wave functions have been developed (most famously that of Newton and Wigner\footnote{The Newton-Wigner (NW) scheme was originally developed in order to have a position operator in relativistic quantum theory, which leaves the positive energy property of a positive energy wave functions invariant. However, the price to pay turns out to be unacceptable: It leads to a deviation from Born's rule, a probability density which does not satisfy a continuity equation with respect to some probability current, the successful minimal coupling to an electromagnetic field does not work in the NW-representation and it violates Lorentz invariance in the sense that a NW-localized state in some Lorentz frame is not NW-localized in any other frame (see \cite{MyBook} and references therein). Nonetheless, the eigenstates of the NW-Operator are (in ordinary position representation) extremely localized Bessel-type functions of positive energy, which, beyond the characteristic length scale of the particle under consideration, virtually look like delta functions. } \cite{NewWig}, but see also e.g. \cite{philips1964lorentz, bracken1, bracken2} and the discussion of these schemes in \cite{MyBook}) which illustrate very nicely that such wave functions can be virtually zero already a few Compton wavelengths or even less away from their center. Moreover, it is straightforward to argue that an electron, for instance, which has interacted with an apparatus or, more generally, with its environment will have an extremely well localized wave function. Such localization processes are well understood in the context of decoherence theory (see, e.g., \cite{hornberger} and references therein).

Since wave functions in quantum theory are the amplitudes of a probability measure, this means that tails can be neglected for all practical purposes. Just as we disregard predictions of negligible probability in thermodynamics (such as rocks suddenly flying up instead of down due to a fluctuation in the thermal velocities of their molecules), we must of course do the same in quantum mechanics, so that we can safely assume wave functions to be compactly supported when making empirical predictions.

But its good to  be aware of the fact that in textbook quantum mechanics the probability interpretation of wave functions is a postulate and there is no statistical analysis (such as Boltzmann's statistical analysis of classical mechanics) to justify it. In Bohmian mechanics, on the other hand, a theory that describes matter as composed of literal particles that always have a position and whose motion is guided by their quantum mechanical wave function, such a statistical analysis can be performed \cite{quantequi, DetlShellyNinoBook, Detlef}. That way, by analyzing the Bohmian equations of motion for measurement-like situations, the quantum probabilities can be derived as predictions for associated (typical) empirical relative frequencies by proving a law of large numbers. And the crucial assumption that goes into a proof of the law of large numbers (and thus, from the Bohmian point of view, establishes the quantum probabilities that are so successful for predictions) is that incredibly improbable events will not happen with empirical certainty (sometimes called Cournot's principle).

When the meaning and status of probabilities is less clear, the issue of infinite tails may be more problematic. This becomes particularly obvious in the Many-Worlds interpretation (MWI), where even the smallest probability events will (at least in a measurement context) actually be realized in some world. However one may interpret the quantum probabilities in MWI and however one may define its ontological content, one probably cannot avoid the fact that there are real worlds in which the infinite tails of positive energy wave functions are empirically relevant (see \cite{DustinAndMe} for details and a remarkable example). 

\vspace{.5cm}

\noindent \textbf{Transformations:} The nonlocal nature of relativistic positive energy wave functions seems to be physically more interesting than infinite tails. Local transformations of relativistic positive energy wave functions necessarily lead to contributions of negative energy states in the resulting state. Moreover, even nonlocal transformations must be extremely special in order to rescue the positive energy property since the values of a relativistic positive energy wave functions in any neighborhood already determines the whole function. And so it can be assumed that at the level of description of one-particle (or N-particle) wave functions, transitions between negative and positive spectrum necessarily occur in physical processes (free time evolution is perhaps the only non-trivial and obvious transformation that is special enough to leave spectral subspaces invariant).

Let us now commit ourselves to the special choice of the Dirac equation, which is the basis for the description of fermions and thus for the description of matter (electrons, quarks, etc.). However, the 
level on which the theory of fermions is not only empirically adequate but impressively successful in its predictions (antimatter, pair creation, Lamb shift etc.) is not that of one-particle or N-particle solutions of the Dirac equation but that of the associated quantum field theory (QFT), in case of the Dirac equation (external field) quantum electrodynamics (QED). This theory can be developed starting from the Dirac equation by second quantization (or more picturesquely from the Dirac sea picture) by allowing roughly speaking for a variable number of particles and interpreting negative energy wave functions by the operation of charge conjugation as positive energy wave functions of antiparticles. Transitions between negative and positive energies on the level of solutions of the Dirac equation thereby correspond to particle creation and annihilation processes with certain probabilities when lifted to the level of QED (see, e.g., \cite{greiner, greiner2012quantum, petersdis, thallerbook}).

Thus, the fragility of positive energy wave functions with respect to nontrivial (e.g., local) transformations, discussed above, suggests that interaction (causing such transformations) is intrinsically associated with particle creation and annihilation processes. Of course, it is to be expected that for everyday processes the corresponding probabilities are again negligibly small, only when high energies are involved this is no longer the case.

\section{An Operational Implication}

Now we come to an operational implication of theorem \ref{basres}. It shall be exemplified by a very general framework for describing a spatial detector experiment. The latter may be taken as only a representative of any local measurement (if any measurement device is triggered by a quantum system, the system was detected in the spatial region of the device).  

\subsection{Covariant Detector Formalism}

\noindent \textbf{Quantum Formalism:} First, we assume that the probability that a detector covering a given spatial region is triggered by a quantum system at a given time (in the lab frame) can be expressed and calculated by the quantum formalism. This means that the click probability in the lab frame is given by an expression of the form
\be \label{qmForm}
\prob{\mathscr{D}_{(0,\Delta)}}=\braket{\psi}{D_{(0,\Delta)}\,\psi}
\ee
Here, $\mathscr{D}_{(0,\Delta)}$ represents the event that a detector covering detector region $\Delta\subset\R^3$ is triggered at lab-time $t=0$, the `probability operator' ${D}_{(0,\Delta)}$ (sometimes called `effect') has the property  $0\leq {D}_{(0,\Delta)}\leq 1$ and shall be an operator in the Heisenberg picture which acts on the Hilbert space of the measured system $\mathcal{H}$ and $\psi\in\mathcal{H}$ is the initial (pure\footnote{We might also work with the more general expression 
 \be
\mathbb{P}^{\, \rho}\left(\mathscr{D}_{(0,\Delta)}\right)=\tra{}{D_{(0,\Delta)}\,\rho}
 \ee
 where $\rho$ is the initial density  matrix, which need not be a pure state. However, since mixed states can always be expressed by (convex) linear combinations of pure states, we can build the following analysis on expression \eqref{qmForm} without loss of generality.}) state. For instance, in the standard ideal measurement scheme of textbooks $D_{(0,\Delta)}$ would be a projection but more generally and more adequate for realistic measurements it is an element of a (not necessarily projective) POVM. 
\vspace{.5cm}

\noindent \textbf{Space-Time Translations:} There is a unitary representation of space-time translations acting on $\mathcal{H}$ which has spectral representation\footnote{The fact that $U(x)$ can be written in this form is of course well known for concrete models of (relativistic) quantum theory and is ensured more generally by an immediate generalization of Stone’s theorem from unitary strongly continuous representations of one parameter groups to unitary strongly continuous representations of general locally compact abelian groups, which is sometimes called the SNAG-theorem (according to Stone, Naimark, Ambrose and Godement) \cite{mackeysnag}.} 
\be 
U(x)=e^{i\widehat{P}x}=\int e^{ipx}d^4E(p)
\ee
Here $\widehat{P}x=\widehat{P}_{\mu}x^{\mu}$ an $px=p_{\mu}x^{\mu}$ are the Minkowski scalar products, $E$ a PVM on $\R^4$ acting on $\mathcal{H}$, the PVM of the energy-momentum operator $\widehat{P}^{\mu}=\int p^{\mu} \, d^4E(p)$ which can be identified as the infinitesimal generator of space-time translations.

 \vspace{.5cm}

\noindent \textbf{Space-Time Translation Covariance:} We assume that space-time translations act naturally on the operators $D_{(0,\Delta)}$: If $x=(s,\vector{a})\in\R^4$, the probability that a detector covering $\Delta+\vector{a}$ is triggered at time $t=s$ in the laboratory frame is given by 
\be 
\prob{\mathscr{D}_{(s,\Delta+\vector{a})}}=\braket{\psi}{D_{(0,\Delta)+x}\,\psi}
\ee
where $D_{(0,\Delta)+x}=U(x)D_{(0,\Delta)}U^{-1}(x)\equiv {D}_{(s,\Delta+\vector{a})}$.

\subsubsection{Additional Assumptions}\label{AdAsSect}

To obtain the desired operational result (theorem \ref{malathm} below), the covariant detector formalism must satisfy some additional assumptions:  \vspace{.5cm}

\noindent \textbf{Spectrum Condition:} We assume that the generator $P_{\mu}$ of space time translations (the energy-momentum operator) has its spectrum in the closed forward light cone: $\sigma(P_{\mu})\subset \overline{V}_+=\{p\in\R^4\mid p_{\mu}p^{\mu}\geq 0 ,\,p_0\geq 0\}$ (see section \ref{basthmsect}).
  \vspace{.5cm}

\noindent \textbf{Additivity:} Now comes a very special assumption. We assume that for $\Delta\cap\Delta'=\emptyset$ and all $\psi\in\mathcal{H}$ there is a joint distribution of the events $\Dtd$ and $\Dtdp$ such that 
\be \label{addi}
\prob{\Dtd \vee \Dtdp}=\prob{\Dtd}+\prob{\Dtdp}
\ee 
This assumption is not justified for general quantum systems; rather, it corresponds to a selection of very special quantum systems for which it appears to be a reasonable assumption. Indeed, the existence of a joint distribution alone only implies (see \cite{MyBook})
\be 
\prob{\Dtd \vee \Dtdp}=\prob{\Dtd}+\prob{\Dtdp} - \prob{\Dtd \wedge \Dtdp}
\ee 
Therefore, equation \eqref{addi} is equivalent to the requirement
\be 
 \prob{\Dtd \wedge \Dtdp}=0 
 \ee 
 i.e., it expresses the requirement that distant detectors cannot be triggered at the same time, given $\psi$ is the initial state. Making this assumption for all $\psi\in\mathcal{H}$ seems to be justified if $\mathcal{H}$ is a Hilbert space of one particle wave functions, which might be taken to be also a subspace of a larger Hilbert space like the one particle sector of Fock space. If we set now
\be \label{addicond}
 \quad D_{(t,\Delta)\cup(t,\Delta')}:= D_{(t,\Delta)}+D_{(t,\Delta')}
\ee 
we thus obtain $\, \prob{\Dtd\vee\Dtdp}=\braket{\psi}{D_{(t,\Delta)\cup(t,\Delta')}\,\psi}\,$.

Additivity is actually not an independent assumption but rather a motivation for its relativistic generalization, causal additivity, which includes additivity as a special case:
\vspace{.5cm}

\noindent \textbf{Causal Additivity:} In a relativistic theory, the natural generalization of additivity is the following: whenever $(t,\Delta)$ and $(t',\Delta')$ are spacelike separated
\be 
\prob{\Dtd \vee \D_{(t',\Delta')}}=\prob{\Dtd}+\prob{\D_{(t',\Delta')}}
\ee
This condition is equivalent to the exclusion of joint detector clicks of two distant detectors at spacelike separation, i.e., 
\be 
\prob{\Dtd \wedge \D_{(t',\Delta')}}=0 
\ee 
By setting $D_{(t,\Delta)\cup(t',\Delta')}\equiv D_{(t,\Delta)}+D_{(t',\Delta')}$ we thus obtain 
 $\prob{\Dtd\vee\D_{(t',\Delta')}}=\braket{\psi}{D_{(t,\Delta)\cup(t',\Delta')}\,\psi}\,$. This condition can be appropriately called \textit{causal additivity} \cite{MyBook}. 
\vspace{.5cm}

\noindent \textbf{Local Commutativity:} The last (relativity inspired) condition we need is the well known condition of local commutativity: whenever $(t,\Delta)$ and $(t',\Delta')$ are spacelike separated
\be 
\left[D_{(t,\Delta)},D_{(t',\Delta')}\right]=0
\ee
This condition is usually demanded to exclude the possibility to use quantum nonlocality in order to send signals faster than light (L\"uders theorem). For a detailed discussion of this condition and further physical motivations see Chapter 3 of \cite{MyBook}.

\subsection{A No-Go Theorem}

Theorem \ref{basres} now implies the following result\footnote{Theorem 2 goes back to a theorem proved in its first version by Schlieder \cite{schliederkaus} and then gradually refined by Jankewitz \cite{jancewicz}, Malament \cite{malament} and Halvorson and Clifton \cite{halvorsonno}, often known as Malament's Theorem.}.
\vspace{-.5cm}
\begin{thm} \label{malathm}
A (non trivial) covariant detector formalism which satisfies the spectrum condition, local commutativity and causal additivity does not exist.
\end{thm} \vspace{-.5cm}

The proof can be found in \cite{MyBook} (theorem 4.25). Roughly speaking, it applies space-time translations to various detector arrangements\footnote{To be precise, the proof uses the obvious generalization of the causal additivity condition to arrangements with more than two detectors, but we skip that here for simplicity.} and thus shows that all click probabilities have an upper bound which can be made inductively arbitrarily small (in this sense, `non-trivial' in theorem \ref{malathm} means `with non-vanishing click probabilities'). The crucial step uses theorem \ref{basres} by applying it to functions $f$ of the form $f(x)=\braket{\varphi}{U(x)\,\psi}$.

\subsection{Discussion} \label{disscet2}

Since there are detectors in the world which can be triggered by quantum systems\footnote{As mentioned above, detector experiments are in this analysis only a representative of practically any quantum measurement (the measured system is detected in the laboratory). One might even argue somewhat drastically that our perception of matter is of this kind in the first place (given the measurement problem has been solved): When I see the table in front of me, I detect the position of a quantum system, given by a huge cluster of atoms, which together form a table.}, theorem \ref{malathm} requires an explanation. One might question any of its assumptions, but of course the assumption of causal additivity is most questionable. Moreover, the discussion of the fragility of the positive energy property of relativistic wave functions with respect to nontrivial transformations together with the observation that spectral transitions of Dirac wave functions correspond to particle creation an annihilation processes in QED in section \ref{dissect1} motivates also a closer look at the spectrum condition. 

Thus, we shall not question here the assumption that the statistics of detector clicks can be predicted by a covariant detector formalism and that local commutativity is true. So according to theorem \ref{malathm}, either the spectrum condition or causal additivity must be violated. Fortunately, these two options naturally complement each other. According to quantum theory, each measurement is associated with a state transformation\footnote{\label{statrafoot}For instance, the probability operator $D$ associated with a triggered detector (for simplicity we suppress the subscript $(t,\Delta)$ here) can be associated with a state transformation operator $\mathcal{R}$ so that an initial state $\psi$ transforms according to $\psi\mapsto\frac{\mathcal{R\,\psi}}{\|\mathcal{R\,\psi}\|}$ and $D=\mathcal{R}^{\dagger}\mathcal{R}$ (for ideal measurements of textbooks, $D$ and $\mathcal{R}$ would be one and the same projection operator, which corresponds to the projection postulate). For more general measurements which cannot be described on the level of pure states, the state transformation is associated with a set $\{\mathcal{R}_k\}$ of linear operators, so that an initial density matrix $\rho$ transforms according to $\rho\mapsto\frac{\sum_k\mathcal{R}^{\dagger}_k\,\rho\,\mathcal{R}_k}{\tr \sum_k\mathcal{R}^{\dagger}_k\,\rho\,\mathcal{R}_k}$ (Kraus representation). See \cite{MyBook} for a detailed development of the general measurement formalism.} (also one with a negative outcome, like a switched on detector which was not (yet) triggered). And as argued in section \ref{dissect1}, most state transformations (in particularly if caused by a localized measuring device) cause spectral transitions on the one- or N-particle level of description, which in turn correspond for fermions to pair creation processes with certain probabilities in the associated QFT. This suggests to expect, for the quantum mechanical description of detector experiments, a violation of the spectrum condition at the level of Dirac wave functions and, when these processes have been lifted to the level of QED, corresponding transitions between the particle number sectors of the fermionic Fock space (while the spectrum condition is rescued in QED by charge conjugation of negative energy states). And the latter immediately destroys any basis for expecting causal additivity to hold in certain situations (one-particle initial states).

To see this, recall that causal additivity corresponds to the assumption that two distant detectors cannot be triggered at spacelike separation and its violation is therefore equivalent to the condition 
\be \label{viocausadd}
\prob{\Dtd \wedge \D_{(t',\Delta')}}>0 
\ee 
for spacelike separated $(t,\Delta)$ and $(t',\Delta')$. For initial states $\psi$ in the one-particle sector of Fock space\footnote{Theorem \ref{malathm} can be generalized to an analogous assertion corresponding to any $N-$particle sector of Fock space: initial states for which it can be perfectly excluded that more than $N$ detectors (for any $N\in\N$) are triggered at spacelike separation do not exist under the assumptions (see corollary 4.27 in \cite{MyBook}).}  this appears to be against the spirit of relativity (a particle moving faster than light to trigger two detectors at spacelike separation). However, if the state transformations associated with such measurements do not leave the one particle sector of Fock space invariant, this violation appears quite natural. For instance, the state transformation caused by the potential of a switched on detector can create a particle by which this detector is being triggered. Since the state transformation associated with a probability operator $D$ is encoded in a linear operator $\mathcal{R}$ so that $\prob{\cdot}=\braket{\psi}{D\,\psi}=\|\mathcal{R}\,\psi\|^2$ (see footnote \ref{statrafoot}), the state transformation, in a sense, enters into the probabilities: even if $\psi$ was a state of a single particle, the predicted statistics can be statistics of many-particles if $\mathcal{R}$ does not leave the one-particle sector of Fock space invariant. Such operators are also well known in connection with observable quantities; the PVM of the local charge density operator in QED, for example, has this property \cite{RodiPosPosOps}. 

This fits very well with a well-known result from the more abstract framework of axiomatic or algebraic quantum field theory (AQFT), the \textit{Reeh-Schlieder theorem} (see, e.g., \cite{witten2018notes} for a comprehensive discussion), which can be also derived from a generalization of theorem \ref{basres} (see \cite{MyBook}). The Reeh-Schlieder theorem implies (under the assumptions of AQFT) that the click probability of a local detector cannot be (exactly) zero even if the initial state is the vacuum state.

To conclude this discussion, note that the fact that causal additivity must be violated says nothing about the magnitude of this violation. The probabilities in \eqref{viocausadd} expressing this violation can be negligibly small, though not precisely zero. If no high energies are involved, negligibly small probabilities \eqref{viocausadd} are of course to be expected for one particle initial states $\psi$.

\section{Towards a Spatial Distribution}

The way the probability operators $D_{(t,\Delta)}$ were defined above, they belong in the first place to a two element POVM $\left\{D_{(t,\Delta)}\, , \, \mathds{1}_{\mathcal{H}}-D_{(t,\Delta)}\right\}$ associated with two possible outcomes (say `click $\equiv 1$' and `no click $\equiv 0$'), which is the minimal structure to describe a detector experiment. However, one has in mind a more general structure, namely a general spatial distribution of a quantum system, which agrees with the click-probabilities given by this POVM for the detector regions.

Theorem \ref{malathm} now also proofs the non-existence of a relativistically satisfying more general spatial POVM on physical space $\R^3$ (instead of $\{0\, ,1\}$) under its assumptions (spectrum condition etc.). To see this, one can simply replace the meaning of the detector regions $\Delta\subset\R^3$, with arbitrary Borel sets $\Delta\subset\R^3$ of physical space. So consider now a spatial POVM in the Heisenberg picture acting on the considered Hilbert space, formed (at a fixed lab-time $t$) by positive operators $D_{(t,\Delta)}$ with $\Delta$ varying in the (measurable) subsets of $\R^3$. As a POVM, it must be additive, i.e., $ D_{(t,\Delta)\cup(t,\Delta')}= D_{(t,\Delta)}+D_{(t,\Delta')}$ for all $\Delta\cap\Delta'=\emptyset$ and normalized, i.e., $\int_{\R^3}D_{(t,d^3x)}=\mathds{1}_{\mathcal{H}}$ is the identity operator (normalization does not play any role for the present considerations). The additivity of such a POVM directly corresponds to the additivity condition \eqref{addicond} above and expressing it in terms of probabilities (i.e., $\prob{\mathscr{D}_{(t,\Delta)}}=\braket{\psi}{D_{(t,\Delta)}\,\psi}$ etc.) yields $\prob{\Dtd \vee \Dtdp}=\prob{\Dtd}+\prob{\Dtdp}$ and thus again $ \prob{\Dtd \wedge \Dtdp}=0$ now for all disjoint spatial Borel sets $\Delta\cap\Delta'=\emptyset$. Calling the event $\Dtd$ sloppily `the system is localized in $\Delta$' we can thus phrase the additivity condition by `the system cannot be localized in two disjoint regions at the same time' (a condition which is clearly false for, say, a two particle system). Causal additivity in this sense means that `the system cannot be localized in two spacelike separated regions', a condition which is the natural relativistic generalization of additivity.

Theorem \ref{malathm}, reformulated with respect to a spatial POVM, then says that such POVM does not exist under the assumptions and thereby a corresponding probability distribution on physical space $\R^3$ does not exist. But what about the $|\psi(\vector{x})|^2-$distribution, which lays the foundation for the predictive success of quantum theory? If we want to express this distribution by a POVM, say for a positive energy solution $\psi$ of the Dirac equation, there are two options at hand: One can use the indicator functions $\chi_{\Delta}(\vector{x})$ of (measurable) spatial subsets $\Delta\subset\R^3$ (the PVM of the standard position operator) or their projection\footnote{The projection operator $P_+$ onto the positive energy subspace of the Hilbert space of solutions of the Dirac equation can be written as $P_+=\frac{1}{2}\left(\mathds{1}_{\mathcal{H}}+\frac{\vector{\alpha}\cdot\vector{p}+\beta m}{\sqrt{\vector{p}^2+m^2}}\right)$, with the usual meaning of the symbols, see, e.g., \cite{thallerbook}.} $P_+\,\chi_{\Delta}(\vector{x})\,P_+$ onto the positive energy subspace of the associated Hilbert space $\mathcal{H}=L^2(\R^3,d^3x)\otimes\C^4$ (both of which form a POVM on $\R^3)$, since for $\psi\in\mathcal{H}_+=P_+\mathcal{H}$ we have the $|\psi|^2-$weight of $\Delta$
\be 
\prob{\Delta}=\int_{\Delta}|\psi(\vector{x})|^2d^3x=\braket{\psi}{\chi_{\Delta}(\vector{x})\,\psi}=\braket{\psi}{P_+\,\chi_{\Delta}(\vector{x})\,P_+\,\psi}
\ee
However, both of these POVMs violate assumptions of theorem \ref{malathm}: multiplication of a positive energy wave function by the indicator functions obviously violates the spectrum condition by radically cutting off everything from the wave function outside of $\Delta$ which yields massive contributions from negative energy eigenstates (observe that an infinite potential well, i.e. an infinite amount of energy would be necessary to realize this operation physically) while their projection onto the positive energy subspace violates local commutativity. For the latter fact, theorem \ref{malathm} can be taken as a proof, but one may also prove it by direct calculation. 

Nonetheless, the probability distribution given by $\prob{\Delta}=\int_{\Delta}|\psi(\vector{x})|^2\, d^3x$ is well defined for positive energy states $\psi$, as long as we do not consider state transformations. A state transformation does not occur if a particle `is there' (say in Bohmian mechanics) but occurs upon measurement\footnote{\label{footstatra} In particular, if $D$ is an element of a POVM, the state transformation upon the associated measurement result is of the form $\psi\mapsto\mathcal{R}\,\psi=U\sqrt{D}\,\psi$ (or a generalization of this formula, if the measurement transforms pure states to mixed states), where $U$ is a partial isometry. If $U=\mathds{1}_{\mathcal{H}}$ and $\sqrt{D}=D=D^2$ is a projection, we recover the projection postulate for ideal measurements. See \cite{MyBook} for details, see also footnote \ref{statrafoot}.}

\section{Particle Ontology}

In view of the previous discussion it is clear, in principle, that the mentioned results do not pose a problem for a quantum theory with a particle ontology, provided it is able to describe particle creation and annihilation (which, of course, it should be for other reasons as well, if it is to reproduce the results of empirically successful relativistic quantum field theories). 

There are several proposals for generalizing non-relativistic Bohmian mechanics to relativistic\footnote{In this context, relativistic QFT refers to QFT with particle creation and annihilation, based on a realtivistic wave equation. The question of full Lorentz invariance is another question which is not treated in this work. Both regularization of the OFT and a description of $N$ particles with nonlocal dynamics pose challenges for a fully Lorenzt invariant description; for treatment of the second point, cf. \cite{beck2020wavefunctions, HBDM, bmlorentz}} QFT \cite{undivuniv, wardcolin, bmdiracsea, bmqft1, bmqft4, bmqft2, wardqft2, wardqft1}. The most elaborated of of these approaches is the so called \textit{Bell type QFT} \cite{bmqft1, bmqft4, bmqft2}, which can be described in a very simplified way as follows: The configuration space is the collection of the configuration spaces (sectors) for each possible particle number\footnote{Details of treatment of identical particles, different particle species etc. are skipped here.} (and antiparticle number) and each sector is associated with a wave function (non-normalized and possibly zero) from the $N-$particle sector of the corresponding Fock space. The actual Bohmian configuration lives in a definite sector ($N$ particles) at each instant and its distribution there is a $|\psi_N|^2-$distribution, $\psi_N$ being the respective sector wave function. In the absence of jumps to other sectors (see below) the actual configuration is deterministically guided by the corresponding sector wave function through a Bohmian guiding equation (for the guiding equation of Dirac theory, see, e.g., \cite{bmqft2}). An additional stochastic jump law provides us with probabilities for where and when particles may be created and/or annihilated (the jump process is driven by the interaction part of the second quantized Hamiltonian). For a given QFT (like regularized QED), these laws define a Markov process on the configuration space consisting of deterministic motion in an actual sector interrupted by stochastic jumps between the sectors, from which the empirical predictions (like cross sections, Lamb shift etc.) of this theory can be derived. 

So the crucial question is how the $|\psi|^2-$distribution of Bohmian configurations fits with the absence of such a distribution for position measurements. In the non-relativistic case, the Bohmian positions of course agree with the results of (good) position measurements, at least to a good degree of accuracy\footnote{Empirical distributions of real world measurements must always minimally deviate from this prediction because measurements are never perfect but are always subject to certain errors with certain probabilities (such errors arise even at the fundamental level due to the quantum mechanical nature of measuring devices \cite{MyBook}). Implementing these uncertainties into the measurement scheme leads to an \textit{approximate measurement POVM}, where the indicator functions of the standard position PVM are convoluted with an additional error distribution, cf. \cite{MyBook}.}. In a relativistic Bohmian QED, this should be the case as well. However, there is a notable difference: the state transformation of a position measurement now not only localizes the wave function of a measured system (or suppresses it in regions where the measurement result is negative) in its actual particle sector, but also generates transitions in the particle number of the measured system with certain probabilities. The presence of a measuring apparatus can thus change the configuration of a measured system by changing the (actual sector of) configuration space. This in turn changes the probabilities of outcomes of the measurement (even if this change will be negligible, if no too large energies are involved). 

Therefore, one should expect that the POVM describing a Bohmian position measurement deviates to some degree from the POVM describing the actual distribution of Bohmain positions. While an electron's Bohmian position, for instance, is $|\psi|^2-$distributed, where $\psi$ is a one particle wave function of positive energy, we should not expect the statistics of its position measurement to be given (precisely) by the corresponding POVM $\left\{ P_+\,\chi_{\Delta}(\vector{x})\, P_+\right\}$  (possibly lifted to Fock space), which does not commute locally\footnote{Observe that the motivation to require local commutativity is to exclude the possibility that a nonlocal state transformation upon measurement (see footnotes \ref{statrafoot} and \ref{footstatra}) can be used to send superluminal signals (see chapter 3 of \cite{MyBook} for a detailed analysis). For probability operators which are not associated with such state transformations, there is no justification for such a requirement.}, but by an operator which includes possible transitions in the particle number due to the intervention of the measuring device. A generic option for an operator describing position measurements\footnote{One might suggest to use the standard position PVM given by the indicator functions $\chi_{\Delta}(\vector{x})$. However, its violation of the spectrum condition is too strong so that any attempt to directly lift it to Fock space by second quantization will fail, because, roughly speaking, its action on Fock space would create infinitely many pairs, as can be estimated, e.g., from its Foldy-Wouthuysen representation (cf. \cite{thallerbook}). } for fermions would be the PVM of the local charge density operator \cite{RodiPosPosOps}, which commutes locally but does not leave the one particle sector of Fock space invariant and hence violates causal additivity. When looking at a concrete position measurement, the question of the associated POVM of course depends on the theoretical modeling of the details of the measurement interaction (detector model).

It is interesting to note that another Bohmian dynamics corresponding directly to the statistics given by the PVM of the local charge density operator can also be defined quite naturally for fermionic Bell type QFT as shown in \cite{RodiPosPosOps}. This theory is empirically equivalent to the fermionic Bell type QFT sketched above (as both are empirically equivalent to regularized QED of textbooks), but they are not equivalent on the ontological level. While in the absence of interaction in the latter case there is no particle creation and annihilation, in the former case configurations can jump between the sectors even under the free time evolution. 

\vspace{1.5cm}

\noindent \textit{Acknowledgements:} I wish to thank Dustin Lazarovici for many productive discussions and for valuable comments on an earlier version of this Article. I would also like to thank Roderich Tumulka for helpful discussions on this topic in the past. And I wish to express my deep gratitude to my teacher and friend Detlef Dürr, for the unique way he taught us and for his unique kindness and warmth.

\newpage 

\bibliographystyle{acm}
\bibliography{bibl}

\def\polhk#1{\setbox0=\hbox{#1}{\ooalign{\hidewidth
  \lower1.5ex\hbox{`}\hidewidth\crcr\unhbox0}}}
\begin{thebibliography}{10}

\bibitem{beck2020wavefunctions}
{\sc Beck, C.}
\newblock {Wavefunctions and Minkowski Space-Time - On the Reconciliation of
  Quantum Theory with Special Relativity}.
\newblock {\em arXiv preprint arXiv:2009.00440\/} (2020).

\bibitem{MyBook}
{\sc Beck, C.}
\newblock {\em Local Quantum Measurement and Relativity}.
\newblock Springer, 2021.

\bibitem{undivuniv}
{\sc Bohm, D., and Hiley, B.~J.}
\newblock {\em {The undivided universe}}.
\newblock Routledge, London, 1995.
\newblock An ontological interpretation of quantum theory.

\bibitem{borcherslemma}
{\sc Borchers, H.-J.}
\newblock {A remark on a theorem of B. Misra}.
\newblock {\em Communications in Mathematical Physics 4}, 5 (1967), 315--323.

\bibitem{bracken1}
{\sc Bracken, A., Flohr, J., and Melloy, G.}
\newblock {Time-evolution of highly localized positive-energy states of the
  free Dirac electron}.
\newblock In {\em Proceedings of the Royal Society of London A: Mathematical,
  Physical and Engineering Sciences\/} (2005), vol.~461, The Royal Society,
  pp.~3633--3645.

\bibitem{bracken2}
{\sc Bracken, A., and Melloy, G.}
\newblock {Localizing the relativistic electron}.
\newblock {\em Journal of Physics A: Mathematical and General 32}, 34 (1999),
  6127.

\bibitem{wardcolin}
{\sc Colin, S., and Struyve, W.}
\newblock {A Dirac sea pilot-wave model for quantum field theory}.
\newblock {\em Journal of Physics A: Mathematical and Theoretical 40}, 26
  (2007), 7309.

\bibitem{bmdiracsea}
{\sc Deckert, D.-A., Esfeld, M., and Oldofredi, A.}
\newblock {A persistent particle ontology for QFT in terms of the Dirac sea}.
\newblock {\em arXiv preprint arXiv:1608.06141\/} (2016).

\bibitem{HBDM}
{\sc D\"urr, D., Goldstein, S., M\"unch-Berndl, K., and Zangh\`\i{}, N.}
\newblock {Hypersurface Bohm-Dirac models}.
\newblock {\em Phys. Rev. A 60}, 4 (Oct 1999), 2729--2736.

\bibitem{bmlorentz}
{\sc D{\"u}rr, D., Goldstein, S., Norsen, T., Struyve, W., and Zangh{\`\i}, N.}
\newblock {Can Bohmian mechanics be made relativistic?}
\newblock {\em Proc. R. Soc. A 470}, 2162 (2014), 20130699.

\bibitem{bmqft1}
{\sc D{\"u}rr, D., Goldstein, S., Tumulka, R., and Zanghi, N.}
\newblock {Trajectories and particle creation and annihilation in quantum field
  theory}.
\newblock {\em Journal of Physics A: Mathematical and General 36}, 14 (2003),
  4143.

\bibitem{bmqft4}
{\sc D{\"u}rr, D., Goldstein, S., Tumulka, R., and Zanghi, N.}
\newblock {Bohmian mechanics and quantum field theory}.
\newblock {\em Physical Review Letters 93}, 9 (2004), 090402.

\bibitem{bmqft2}
{\sc D{\"u}rr, D., Goldstein, S., Tumulka, R., and Zanghi, N.}
\newblock {Bell-type quantum field theories}.
\newblock {\em Journal of Physics A: Mathematical and General 38}, 4 (2005),
  R1.

\bibitem{quantequi}
{\sc D{\"u}rr, D., Goldstein, S., and Zangh{\`{\i}}, N.}
\newblock {Quantum equilibrium and the origin of absolute uncertainty}.
\newblock {\em J. Statist. Phys. 67}, 5-6 (1992), 843--907.

\bibitem{DetlShellyNinoBook}
{\sc D{\"u}rr, D., Goldstein, S., and Zangh{\`\i}, N.}
\newblock {\em {Quantum Physics Without Quantum Philosophy}}.
\newblock Springer Berlin Heidelberg, 2012.

\bibitem{Detlef}
{\sc D\"urr, D., and Teufel, S.}
\newblock {\em {Bohmian Mechanics: The Physics and Mathematics of Quantum
  Theory}}.
\newblock Springer Berlin, 2009.

\bibitem{greiner}
{\sc Greiner, W.}
\newblock {\em {Relativistic quantum mechanics. Wave equations}}.
\newblock Springer, 2000.

\bibitem{greiner2012quantum}
{\sc Greiner, W., M{\"u}ller, B., and Rafelski, J.}
\newblock {\em {Quantum electrodynamics of strong fields: with an introduction
  into modern relativistic quantum mechanics}}.
\newblock Springer Science \& Business Media, 2012.

\bibitem{halvorsonno}
{\sc Halvorson, H., and Clifton, R.}
\newblock {No place for particles in relativistic quantum theories?}
\newblock {\em Philosophy of Science 69}, 1 (2002), 1--28.

\bibitem{hegerfeldt1985}
{\sc Hegerfeldt, G.~C.}
\newblock {Violation of causality in relativistic quantum theory?}
\newblock {\em Physical review letters 54}, 22 (1985), 2395--2398.

\bibitem{hegerfeldt1998a}
{\sc Hegerfeldt, G.~C.}
\newblock {Instantaneous spreading and Einstein causality in quantum theory}.
\newblock {\em arXiv preprint quant-ph/9809030\/} (1998).

\bibitem{hegerfeldt1980}
{\sc Hegerfeldt, G.~C., and Ruijsenaars, S.~N.}
\newblock {Remarks on causality, localization, and spreading of wave packets}.
\newblock {\em Physical Review D 22}, 2 (1980), 377--384.

\bibitem{hornberger}
{\sc Hornberger, K.}
\newblock {Introduction to decoherence theory}.
\newblock In {\em Entanglement and Decoherence}. Springer, 2009, pp.~221--276.

\bibitem{hyperbolPDE}
{\sc Ikawa, M.}
\newblock {\em {Hyperbolic partial differential equations and wave phenomena}},
  vol.~2.
\newblock American Mathematical Soc., 2000.

\bibitem{jancewicz}
{\sc Jancewicz, B.}
\newblock {Operator density current and relativistic localization problem}.
\newblock {\em Journal of Mathematical Physics 18\/} (1977), 2487.

\bibitem{DustinAndMe}
{\sc Lazarovici, D., and Beck, C.}
\newblock {Nonlocal Vacuum Phenomena}.
\newblock {\em In Preparation\/} (2023).

\bibitem{mackeysnag}
{\sc Mackey, G.~W.}
\newblock {Harmonic analysis and unitary group representations: the development
  from 1927 to 1950}.
\newblock {\em Cahiers du S{\'e}minaire d'histoire des math{\'e}matiques 2\/}
  (1992), 13--42.

\bibitem{malament}
{\sc Malament, D.}
\newblock {In Defense Of Dogma: Why There Cannot Be A Relativistic Quantum
  Mechanics Of (Localizable) Particles}.
\newblock {\em Perspectives on Quantum Reality: Non-Relativistic, Relativistic,
  and Field-Theoretic\/} (1966), 1--10.

\bibitem{NewWig}
{\sc Newton, T.~D., and Wigner, E.~P.}
\newblock {Localized States for Elementary Systems}.
\newblock {\em Rev. Mod. Phys. 21\/} (Jul 1949), 400--406.

\bibitem{philips1964lorentz}
{\sc Philips, T.}
\newblock {Lorentz invariant localized states}.
\newblock {\em Physical Review 136}, 3B (1964), B893.

\bibitem{petersdis}
{\sc Pickl, P.}
\newblock {\em {Existence of spontaneous pair creation}}.
\newblock PhD thesis, Ludwig-Maximilians-Universit{\"a}t M{\"u}nchen, 2005.

\bibitem{schliederkaus}
{\sc Schlieder, S.}
\newblock {Zum kausalen Verhalten eines relativistischen
  quantenmechanischenSystems}.
\newblock {\em Quanten und Felder. Braunschweig: Friedrich Vieweg+ Sahn\/}
  (1971), 145--160.

\bibitem{schwablAQM}
{\sc Schwabl, F.}
\newblock {\em {Advanced quantum mechanics}}.
\newblock Springer Science \& Business Media, 2005.

\bibitem{Schweber}
{\sc Schweber, S.~S.}
\newblock {\em {An Introduction to Relativistic Quantum Field Theory}}.
\newblock Dover Publications, June 2005.

\bibitem{wardqft2}
{\sc Struyve, W.}
\newblock {Pilot-wave theory and quantum fields}.
\newblock {\em Reports on Progress in Physics 73}, 10 (2010), 106001.

\bibitem{wardqft1}
{\sc Struyve, W.}
\newblock {Pilot-wave approaches to quantum field theory}.
\newblock In {\em Journal of Physics: Conference Series\/} (2011), vol.~306,
  IOP Publishing, p.~012047.

\bibitem{thallerbook}
{\sc Thaller, B.}
\newblock {The Dirac Equation (Texts and Monographs in Physics)}.
\newblock {\em Springer-Verlag, Berlin 91\/} (1992), 1105--1115.

\bibitem{RodiPosPosOps}
{\sc Tumulka, R.}
\newblock {Positron position operators. I. A natural option}.
\newblock {\em Annals of Physics 443\/} (2022), 168988.

\bibitem{witten2018notes}
{\sc Witten, E.}
\newblock {Notes on Some Entanglement Properties of Quantum Field Theory}.
\newblock {\em arXiv preprint arXiv:1803.04993\/} (2018).

\end{thebibliography}

\end{document}